\documentclass[12pt]{article}
\usepackage{epsfig,amssymb}
\usepackage{latexsym}

\newcommand{\bq}{\begin{equation}}
\newcommand{\eq}{\end{equation}}
\newcommand{\bqa}{\begin{eqnarray}}
\newcommand{\eqa}{\end{eqnarray}}
\newcommand{\ben}{\begin{enumerate}}
\newcommand{\een}{\end{enumerate}}
\newcommand{\bc}{\begin{center}}
\newcommand{\ec}{\end{center}}
\newcommand{\bqb}{\begin{eqnarray*}}
\newcommand{\eqb}{\end{eqnarray*}}

\def\gsim{\gtrsim}
\def\lsim{\lesssim}

\def\pr#1#2#3{ Phys. Rev. ${\bf{#1}}$,#2 (#3)}
\def\prl#1#2#3{ Phys. Rev. Lett. ${\bf{#1}}$,#2 (#3)}
\def\pl#1#2#3{ Phys. Lett. ${\bf{#1}}$,#2 (#3)}
\def\prep#1#2#3{ Phys. Rep. ${\bf{#1}}$,#2 (#3)}
\def\rmp#1#2#3{ Rev. Mod. Phys. ${\bf{#1}}$,#2(#3)}
\def\rpp#1#2#3{Rep. Prog. Phys.   ${\bf{#1}}$,#2 (#3)}
\def\np#1#2#3{ Nucl. Phys. ${\bf{#1}}$,#2 (#3)}

\def\jhep#1#2#3{ JHEP ${\bf{#1}}$,#2 (#3)}
\def\epj#1#2#3{ Eur. Phys. J. ${\bf{#1}}$,#2 (#3)}

\def\cpc#1#2#3{Comput. Phys. Commun. ${\bf{#1}}$,#2 (#3)}
\def\app#1#2#3{Astropart. Phys. ${\bf{#1}}$,#2 (#3)}

\def\ie{{\it i.e. ~}}
\def\eg{{\it e.g. ~}}

\def\etal{{\it et.al.~}}

\global\nulldelimiterspace = 0pt

\def\mw{m_W}

\def\tchi{\tilde \chi}

\begin{document}
\pagenumbering{arabic}
\thispagestyle{empty}
\def\thefootnote{\fnsymbol{footnote}}
\setcounter{footnote}{1}

\begin{flushright}
September 2003, corrected January 2004\\
PM/03-18\\
THES-TP 2003/03-03. \\
hep-ph/0309032.\\

 \end{flushright}
\vspace{2cm}
\begin{center}
{\Large\bf Neutralino-neutralino annihilation to
photon and gluon pairs in MSSM models.\footnote{Programme
d'Actions Int\'egr\'ees Franco-Hellenique, Platon 04100 UM}}
 \vspace{1.5cm}  \\
{\large G.J. Gounaris$^a$, J. Layssac$^b$, P.I. Porfyriadis$^a$
and F.M. Renard$^b$}\\
\vspace{0.2cm}
$^a$Department of Theoretical Physics, Aristotle
University of Thessaloniki,\\
Gr-54124, Thessaloniki, Greece.\\
\vspace{0.2cm}
$^b$Physique
Math\'{e}matique et Th\'{e}orique,
UMR 5825\\
Universit\'{e} Montpellier II,
 F-34095 Montpellier Cedex 5.\\

\vspace*{1.cm}

{\bf Abstract}
\end{center}
 The complete 1-loop computation of the processes
$(\tchi^0_i \tchi^0_j \leftrightarrow\gamma\gamma, ~ gg)$, for any
pair of the four MSSM neutralinos, has been performed
for   an arbitrary  c.m energy. As a first application suitable for Dark
Matter (DM) searches, the neutralino-neutralino annihilation is studied
at relative velocities $v/c\simeq 10^{-3}$ describing
the present DM distribution in the galactic halo,
 and $v/c\simeq 0.5$ determining neutralino relic density contributions.
Numerical results are presented for 31  MSSM benchmark models
indicating considerable sensitivity to
the input  parameters. Numerical codes
are  released, which may be used   for the computation of
the  annihilation of any two neutralinos  to two photons
or two gluons at the aforementioned $v/c$ values.
In the near future, we intend to complement them
with  codes for the inverse   process
$\gamma\gamma \to \tchi^0_i \tchi^0_j$, observable at the
future high energy Linear Colliders.

\vspace{0.5cm}

\def\thefootnote{\arabic{footnote}}
\setcounter{footnote}{0}
\clearpage

The nature of the Dark Matter (DM), which is here assumed to be
 possibly sensitive  only to the usual  gravitational and  weak forces
\cite{Kamio-rep, DMrev}, is still open.
At present, a  most obvious candidate  for such matter in the context of
R-parity conserving Supersymmetric (SUSY) theories,
 should be  the Lightest Supersymmetric Particle (LSP) \cite{LSPDM}.
In  the minimal Supersymmetric models (MSSM),
this is most often identified with the lightest
neutralino $\tchi^0_1$  \cite{mSUGRA-mod, AMSB-mod}. Within MSSM there
exist cases though,  where  the LSP is the  purely
gravitationally interacting  gravitino, while the
lightest  neutralino may then appear as the
next to lightest supersymmetric particle  (NLSP) \cite{GMSB-mod}.\par

In  any case, depending on the MSSM parameters, one or more
neutralinos may sizably contribute to DM  \cite{Kamio-rep, DMrev}.
Such  DM could be investigated though   direct or indirect methods.
Direct detection could  occur  through \eg the observation
of the nucleus  recoil induced by the
$\tchi^0_1$-nucleus  scattering \cite{dirDM}.
Experimental setups are looking for such events \cite{CDMS-exp}, and
there even exist  claims  that a signal may have
already been observed  \cite{DAMA}.

Indirect detection refers to  searches of anomalies in the spectrum
of the photons, neutrinos, positrons or antiprotons, emitted
 from cosmologically nearby   sites,  where the neutralino concentration
 might be enhanced   \cite{Kamio-rep, DMrev}.
Both types of searches are however difficult,
because of  uncertainties in making precise predictions
for the dark matter density distribution in these sites \cite{DM-distribution},
and the strong sensitivity
of  the observable quantities on the   SUSY parameters.\par

However, as noticed several years
ago \cite{nnDM1, nnDM2}, the
process\footnote{$\tchi^0_1$ denotes the lowest neutralino.}
$\tchi^0_1\tchi^0_1 \to\gamma\gamma$ (as well
as $\tchi^0_1\tchi^0_1 \to \gamma Z$),
 should be  particularly interesting
for Dark Matter detection, since  the neutralino pair
annihilation is essentially taking place at rest,
 producing  sharp monochromatic photons. Provided therefore
that the cross section $\sigma(\tchi^0_1\tchi^0_1 \to\gamma\gamma)$
and the neutralino concentration in the galactic halo
are not  very   small, the  signal to be searched for would be
  sharp photons coming from the central region of the Galaxy,
  with energies in the hundred-GeV  region.

  The companion process
$\tchi^0_1\tchi^0_1 \to gg$ may  also be
important in some regions of the parameter space,
where   its sufficiently frequent
occurrence may modify the analyses of the
neutrino and   antiproton signals \cite{Dreesgg}.
Furthermore, if  the second neutralino $\tchi^0_2$ turns out to be
very close to  $\tchi^0_1$,  processes like
$(\tchi^0_1\tchi^0_2 \to\gamma\gamma, ~gg )$
and $(\tchi^0_2\tchi^0_2 \to\gamma\gamma, ~gg )$ may also be
interesting;  particularly the gluonic mode
  may then play a  role in   determining
the relic neutralino density and  the rates of neutrino and antiproton production
\cite{Dreesgg}.

These statements sufficiently motivate  our interest in the general
processes $\tchi^0_i\tchi^0_j \leftrightarrow \gamma\gamma, ~gg$,
( $i,j=1-4$ ). The lowest order  non-vanishing contribution to them
arise  from  one loop diagrams involving
 standard  and   supersymmetric particle exchanges; \ie sleptons, squarks,
gauginos and additional Higgses, which sensitively depend
on the  precise SUSY particle spectrum, couplings and  mixings.
Thus, DM considerations of the processes
$(\tchi^0_i \tchi^0_j\to\gamma\gamma, ~gg)$, and studies of the inverse
processes $\gamma \gamma \to \tchi^0_i \tchi^0_j $ in a Linear Collider (LC),
and $gg  \to \tchi^0_i \tchi^0_j $ at  LHC, may provide very strong
constraints on the various MSSM models.

Previous
 computations have only addressed the diagonal annihilation
$\tchi^0_i\tchi^0_i \to \gamma\gamma,~gg$   of two
identical  neutralinos  at vanishing relative velocity
 \cite{nnDM1, nnDM2, Dreesgg}.
Such  approximations  may   of course be
adequate  for neutralinos annihilating at present
in the galactic halo (or  the Sun or Earth),
where their average relative  velocity  is
 expected to be  $v/c\simeq 10^{-3}$ \cite{Kamio-rep}.
But this  low energy limit  may not be generally
 adequate   for   neutralino relic density computations,
 which is determined by neutralino annihilation at a time
 when their  average relative velocity prevailing in Cosmos
 was  $v/c\simeq 0.5$    \cite{Kamio-rep, various}.

 We have, therefore, made a
complete  one loop computation of
$\tchi^0_i\tchi^0_j \leftrightarrow \gamma\gamma, gg$, in  a general
MSSM model with real parameters, expressing all   helicity amplitudes
in terms of the
Passarino-Veltman functions, for general incoming and
outgoing momenta.
The derived  expressions, which  should be useful in both dark matter
and  LC and LHC studies,   are quite complicated though, so
 that the only efficient way
to present them is through numerical codes.
Details for both types of processes and applications to
  collider physics searches,  will be given in
  a separate extended paper \cite{nn2}.

In the present paper  we restrict to  studies of the
$\tchi^0_i\tchi^0_j \to\gamma\gamma$ and $\tchi^0_i\tchi^0_j \to gg$
 annihilations in the framework of DM searches,
  where the average  relative velocity
  $v_{ij}$ of the two neutralinos  is   small.
   Two numerical codes,
PLATONdml and PLATONdmg are released  \cite{plato},
which  compute the neutralino annihilation cross sections  times
  $v_{ij}$; \ie
$(v_{ij}/c) \cdot \sigma(\tchi^0_i\tchi^0_j \to \gamma\gamma) $ and
 $(v_{ij}/c)\cdot  \sigma (\tchi^0_i\tchi^0_j \to  gg)$ respectively,
at $v_{ij}/c\simeq 10^{-3}$, as it would be
   expected   for galactic halo neutralinos  at present
\cite{nnDM1, nnDM2, Kamio-rep}.
In addition, a third code PLATONgrel, is also released, which  calculates
$(v_{ij}/c)\cdot  \sigma (\tchi^0_i\tchi^0_j \to  gg)$
at $v_{ij}/c\simeq 0.5$, which should be useful for neutralino
relic density computations. The outputs of all  codes  are  in fb.

\begin{figure}[p]
\vspace*{-3cm}
\[
\hspace{-1.cm}\epsfig{file=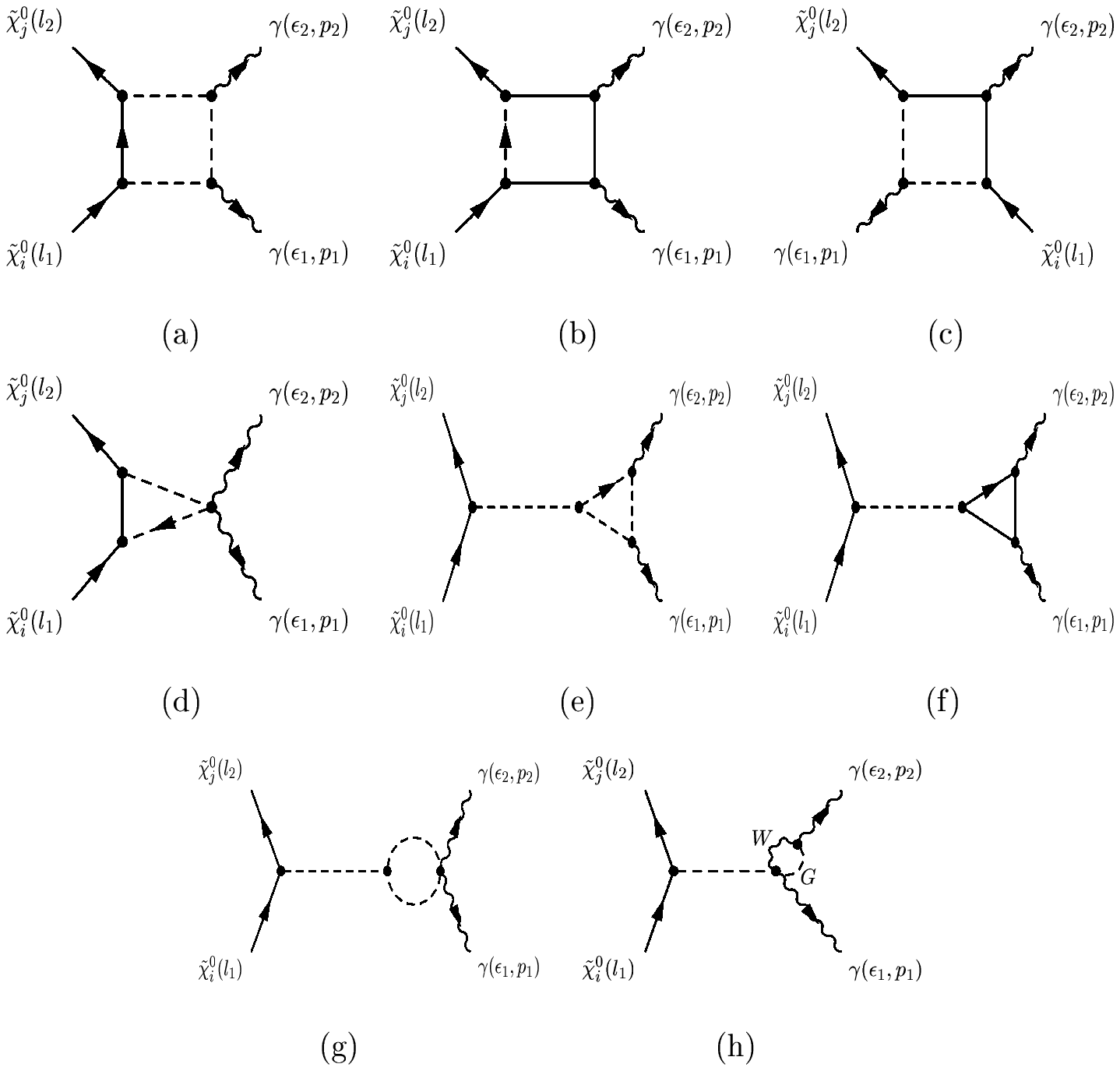,height=13.cm, width=13.cm}
\]
\caption[1]{ Feynman diagrams for $\chi^0_i\chi^0_j \to \gamma\gamma$.
Full internal lines denote fermionic exchanges; while
broken internal lines  denote either scalar or gauge exchanges,
 except in the diagram (h),  where the
 W and Goldstone exchanges are  indicated explicitly.}
\label{Diagrams}
\end{figure}

\vspace{0.5cm}
\noindent
{\it\underline{Procedure}}.~~\\
 The  generic structure of the  one-loop diagrams
 for $\tchi^0_i\tchi^0_j \to \gamma\gamma $
 in the 't-Hooft-Feynman gauge,  is depicted in Fig.\ref{Diagrams},
where the types of  particles running
clockwise inside each  loop   are:
box  (a),   $(fSSS)$, $(fWWW)$, $(fSSW)$, $(fSWW)$, $(fSWS)$,
$(fWWS)$; box (b),   $(Sfff)$, $(Wfff)$;
box (c)  $(SffS)$,  $(WffW)$, $(SffW)$, $(WffS)$;
initial triangle (d),  $(SfS)$, $(WfW)$;
final triangle (e),   $(WWW)$, $(SSS)$, $SWW$, $SSW$;
final triangle  (f), $(fff)$; and bubbles (g) $(SS)$, $(WW)$; (h). $WS$.
By $S$ we denote the  scalar exchanges
(Higgs, Goldstone and  sfermions);
and by $f$  the  fermionic ones  (leptons, quarks and inos).
Bubbles (g) and final
triangles (e) and (f), are connected to the initial $\tchi^0_i \tchi^0_j$ state
through an intermediate $Z$,  or neutral  Higgs or
Goldstone boson $h^0$, $H^0$, $A^0$,  $G^0$. For the
$\tchi^0_i\tchi^0_j \to gg $ process, only the  subset of the diagrams in
Fig.\ref{Diagrams}, involving  quark and squark loops, is needed.
We also note that the $\tchi^0_i\tchi^0_j$-antisymmetry, and the Bose
symmetries of the  $\gamma \gamma$ and
 $gg$ states,  provide   non-trivial tests for the  helicity  amplitudes
of $\tchi^0_i\tchi^0_j \to \gamma\gamma,~ gg $, which we have  checked to be
satisfied  \cite{nn2}.

\vspace{0.5cm}
\noindent
{\it\underline{General features for small relative velocities}}.~~\\
When the relative  velocity $v_{ij}$ among the neutralino pairs   is  small,
 their c.m energy to order in $(v_{ij}/c)^4$, is
\bq
s=(m_i+m_j)^2+m_im_j \Big (\frac{v_{ij}}{c}\Big )^2
\Big [1+  \Big (\frac{v_{ij}}{c}\Big )^2
\Big (\frac{3}{4}-\frac{2 m_i m_j}{(m_i+m_j)^2} \Big ) \Big ]
~~. \label{s-threshold}
\eq
For very small values of\footnote{For  all MSSM-models  we are
aware of, the differential cross section is flat
for   $v_{ij}/c\lsim 0.5 $.} $v_{ij}$,
the angular distribution of
$d\sigma(\tchi^0_i\tchi^0_j \to\gamma\gamma,~  gg)/d\cos\theta$
is   flat, and  $v_{ij} \sigma(\tchi^0_i\tchi^0_j \to\gamma\gamma,~  gg) $
 is  a  smooth function of $v_{ij}^2$.

Since $\sigma(\tchi^0_i\tchi^0_j \to \gamma\gamma, ~ gg)$
are induced by 1-loop processes,
the expected  order of magnitude for very small $v_{ij}/c$  is
\bqa
 v_{ij}\sigma(\tchi^0_i\tchi^0_j \to \gamma\gamma) & \simeq &
 \frac{\alpha^4 c}{M^2_{eff}}
\simeq  \frac{3\times 10^{-32}cm^3sec^{-1}}{M^2_{eff}({\rm TeV})}
 ~~, \nonumber \\
v_{ij} \sigma (\tchi^0_i\tchi^0_j \to gg) & \simeq &
\frac{\alpha^2\alpha_s^2 c}{M^2_{eff}}
\simeq \frac{6\times 10^{-30}cm^3sec^{-1}}{M^2_{eff}({\rm TeV})} ~~,
\label{vij-sigma}
\eqa
where $M_{eff}$ is an effective mass-scale, depending on the
masses and  mixings of the external  neutralinos and the
 particles running inside the loops.
For such   $v_{ij}/c$, the two
neutralinos are predominantly in a $^1S_0$ state, so that only helicity
amplitudes satisfying $\lambda_1=\lambda_2$ and  $\mu_1=\mu_2$ are allowed.

The dominant diagrams   in this case are the
boxes  $(Sfff)$, $(SffS)$ $(fWWW)$,
$(Wfff)$, $(WffW)$, $(fWGW)$, $(GffW)$, $(WffG)$.
Triangle and bubble contributions  are either vanishing or numerically
negligible, except for the
case where the s-channel exchanged neutral Higgs is very  close to the
sum of the two neutralino masses; compare Fig.\ref{Diagrams}e-h.
In such a case, depending on whether  $\tchi^0_i$ and $\tchi^0_j$
 have the same or opposite CP eigenvalues,
the main  contribution arises from the
 exchange of  either $A^0$ or\footnote{The lightest $h^0$
is  below  threshold in all models considered.} $H^0$, respectively.

The  magnitude  of $v_{ij} \sigma $ in (\ref{vij-sigma}),
may  thus  be enhanced, if  relatively light    stops,  sbottoms,
and charginos or staus (for the $\gamma\gamma$ case only)
exist, and  one of them happens to be  almost degenerate
  to one of the  incoming  neutralinos; or if  a
 neutral Higgs,  with  an appropriate  CP eigenvalue,
 is  very close to the c.m. neutralino energy, (compare the
 diagram of  Fig.\ref{Diagrams}e-h).
 But the most important role in determining  the
 $v_{ij}\sigma$-magnitude  seems
 to be played by the nature of the neutralino contents;
 mixture percentages of Wino, Higgsinos or Bino. The
 Wino-type neutralinos usually give  the largest cross sections,
 followed up by the Higgsinos, while  Binos are most often supplying
 very small cross sections. This ordering comes from the dominant role
played by the box diagrams involving $W$ bosons and the
neutralino-chargino-W couplings.
For example, it has been
emphasized  that if the lightest neutralino $\tchi^0_1$
 is Higgsino-like
and close to a chargino, then $M_{eff} \simeq \mw$, leading to
  $v  \sigma(\tchi^0_1\tchi^0_1 \to \gamma\gamma)
  \sim 10^{-28}cm^3sec^{-1} $,  independently of the neutralino mass;
  compare (\ref{vij-sigma}),  \cite{nnDM1, nnDM2}.
We  confirm this result. In fact,  we have numerically
checked   that    our exact 1-loop  computation
reproduces  the  results of \cite{nnDM1, nnDM2}, obtained
through a  non-relativistic treatment
of the annihilation cross section  of two  identical
neutralinos at threshold. It should also be remarked that
 (\ref{vij-sigma}) is consistent with unitarity in our calculations,
provided that the neutralino masses are in the TeV-range or below, \cite{nnunit}.
Thus,  higher order effects can be safely ignored.

\vspace{0.5cm}
\noindent
{\it \underline{Illustrations for various MSSM models}}.~~\\
Using  the numerical codes
PLATONdml,  PLATONdmg  and PLATONgrel, we obtain  the results for
 $v_{ij}  \sigma(\tchi^0_i\tchi^0_j \to \gamma\gamma) $ and
 $v_{ij}   \sigma (\tchi^0_i\tchi^0_j \to  gg)$
 presented in Tables 1, 2 and 3, for an extensive set
 of benchmark SUSY models involving
 real parameters only. These  models are used here just
 in order to indicate the range of possible
 results we could obtain. They all   satisfy  $\mu>0$ and
have been suggested as being roughly
consistent with  present theoretical and experimental constraints
 \cite{Snowmass, Arnowitt, Rosz, CDG, benchmarks2}.
In constructing these  codes,  the exact 1-loop expressions
for $\tchi^0_i\tchi^0_j \to \gamma\gamma,~ gg $,
have been used.

In the same Tables, the grand  scale parameters
of the various  benchmarks MSSM models are also shown.
Conveniently,  these models are divided  into three classes:
\begin{itemize}
\item \underline{ Universal m-SUGRA models.}
Apart from the $\mu$-sign,  there exist four
 additional  GUT scale parameters in these models, namely
 ($m_{1/2}$, $m_0$,  $A_0$, $\tan\beta$). They are
 given in the left part of Table 1. The $SPS$-models in this Table
come from the  Snowmass set  \cite{Snowmass},
 $AD(fg5)_{1,2}$  from Fig.5 of
\cite{Arnowitt}, and $Rosz_2$ from Fig.2 of \cite{Rosz}.
\item \underline{Non-universal m-SUGRA models.}
The grand  scale parameters of theses models are presented
in the upper part of Table 2.  In the   models $AD(fg9)$ (coming
from Fig.9 of \cite{Arnowitt})
and $SPS6$ \cite{Snowmass},  $m_{H_u}$ is allowed to deviate at the grand scale
from the scalar  mass   unification condition;
while in the $CDG$-models,
 the gaugino large scale unification  is violated \cite{CDG}.
\item \underline{GMSB and AMSB models.}
The grand scale parameters for the gauge mediated
symmetry breaking (GMSB) models $(SPS7,~SPS8)$,
and the anomaly mediated symmetry breaking (AMSB)
 models $SPS9$,  are given in Table 3, \cite{Snowmass}.
\end{itemize}

Using a   public code like \eg  SuSpect \cite{suspect}, the electroweak scale
parameters of the above benchmark models may be  calculated
from their grand scale values; while the also needed  total  widths
of the supersymmetric neutral Higgs
particles\footnote{Compare the diagrams in Fig.\ref{Diagrams}e-h.}
$H^0$ and $A^0$,   are calculated
using\footnote{In some non-universal  SUSY models
we are using, an HDECAY version is needed
allowing  independent values for the gaugino parameters
$M_1$ and $M_2$ at the electroweak scale.
We thank Abdelhak Djouadi for providing us with
this code. }  HDECAY \cite{HDECAY}.
These in turn are used
in the in-files of  the three   PLATON   codes.

We next turn to the results for
$v_{ij}  \sigma(\tchi^0_i\tchi^0_j \to \gamma\gamma,~ gg) $ for
the\footnote{Here $(\tchi^0_1, \tchi^0_2)$ are the
lightest and the next to lightest neutralinos.}
$(\tchi^0_1\tchi^0_1)$, $(\tchi^0_1 \tchi^0_2)$ and $(\tchi^0_2\tchi^0_2)$
cases, which appear in Tables 1,2,3.
Of course, the $(\tchi^0_1,\tchi^0_1 \to \gamma\gamma)$ case is
the most interesting one
for Dark Matter detection in the halo, but the other modes
(especially the gluonic ones if they turn out to be appreciable),
 may play an important role for the determination of the value of
the DM relic density
and  the rates of neutrino and antiproton production
\cite{Dreesgg}.
Below we discuss these results, in conjunction
with the electroweak scale masses and couplings of the various models.
\begin{table}[htb]
\begin{center}
{ Table 1: Universal mSUGRA Models: Input parameters are given
at the grand scale,
  and $v_{ij}\sigma (\tchi^0_i \tchi^0_j \to \gamma \gamma, ~gg)$
  predictions are  at $v_{ij}/c \simeq 10^{-3}$ ($v_{ij}/c \simeq 0.5$).  }\\
  \vspace*{0.3cm}
\begin{small}
\begin{tabular}{||c|c|c|c|c||c|c|c|c|c|c||}
\hline \hline
\multicolumn{5}{||c||}{Parameters (Dimensions in $\rm GeV$) }&
\multicolumn{6}{c||}{$v_{ij}\sigma (\tchi^0_i \tchi^0_j
\to \gamma \gamma, ~gg)$ in $10^{-29} cm^3 sec^{-1}$ } \\ \hline
\multicolumn{5}{||c||}{ $\mu>0$ } & \multicolumn{3}{|c|}{ $\gamma \gamma$ }
& \multicolumn{3}{|c||}{ $gg$ }  \\
\hline
Model & $m_0$ & $m_{1/2}$ & $A_0$ & $\tan \beta$
& $ \tchi^0_1 \tchi^0_1 $
&  $ \tchi^0_1 \tchi^0_2  $
& $ \tchi^0_2 \tchi^0_2  $
& $ \tchi^0_1 \tchi^0_1   $
&  $ \tchi^0_1 \tchi^0_2  $
& $ \tchi^0_2 \tchi^0_2  $    \\ \hline
 $SPS5$ & 150 & 300  & -1000 & 5 & 0.234 & 0.199 & 152. & 0.0906 & 0.689 & 8.40 \\
  &  &   &  & &  &  &  & (0.0891) & (1.52) & (9.23) \\
 $SPS1a_1$ & 100 & 250  & -100 & 10 & 0.443 & 0.643 & 144. & 0.460 & 0.517 & 4.40 \\
 &  &   &  & &  &  &  & (0.474) & (0.580) & (8.00) \\
 $SPS1a_2$ & 140 & 350  & -140 & 10 & 0.291 & 0.164 & 139. & 0.212 & 0.205 & 10.6 \\
 &  &   &  & &  &  &  & (0.214) & (0.225) & (28.7) \\
 $SPS1a_3$ & 200 & 500  & -200 & 10 & 0.165 & 0.0444 & 134. & 0.110 & 0.0342 & 31.6 \\
 &  &   &  & &  &  &  & (0.118) & (0.0466) & (54.3) \\
 $SPS3_1$ & 90 & 400  & 0. & 10 & 0.582 & 0.154 & 53.0 & 0.179 & 0.134 & 1553. \\
 &  &   &  & &  &  &  & (0.178) & (0.173) & (25.0) \\
 $SPS3_2$ & 140 & 600  & 0. & 10 & 0.376 & 0.0421 & 114. & 0.098 & 0.0183 & 3.36 \\
 &  &   &  & &  &  &  & (0.102) & (0.0339) & (2.26) \\
 $SPS3_3$ & 190 & 800  & 0. & 10 & 0.176 & 0.0187 & 113. & 0.0636 & 0.00574 & 1.6 \\
 &  &   &  & &  &  &  & (0.0647) & (0.0141) & (1.00) \\
  $SPS2_1$ &1450  & 300  & 0. & 10 & 0.000193 & 0.0666 &150  & 0.0433 & 0.488 & 1.84 \\
 &  &   &  & &  &  &  & (0.0470) & (0.581) & (1.80) \\
 $SPS2_2$ &1750  & 450 & 0. & 10 & 0.00041 & 0.0112 & 134 & 0.0490 &0.162  & 0.611 \\
 &  &   &  & &  &  &  & (0.0490) & (0.161) & (0.606) \\
 $SPS2_3$ & 2050 & 600  & 0. & 10 & 0.000265 & 0.00336 & 125 & 0.0207 &0.0585  &0.281  \\
 &  &   &  & &  &  &  & (0.0214) & (0.0586) & (0.280) \\
$SPS1b$ & 200 & 400  & 0. &30  & 0.197  & 0.0651 & 120 & 0.337 &1.10  & 6.06 \\
 &  &   &  & &  &  &  & (0.350) & (1.40) & (4.17) \\
 $AD(fg5)_1$ &220  & 400  & 0. & 40 & 0.238 & 0.0635  & 121  &0.574  & 19.1 &4.03  \\
 &  &   &  & &  &  &  & (0.620) & (90.6) & (3.97) \\
 $AD(fg5)_2$  & 400 & 900  & 0. & 40 & 0.0524 & 0.0153  & 113 &0.0516  & 10.4 &0.814  \\
 &  &   &  & &  &  &  & (0.0528) & (0.823) & (0.828) \\
 $SPS4$ & 400 & 300  & 0. & 50 &0.0424  & 0.0685 & 130 & 5.69 & 14.7 & 7.5 \\
 &  &   &  & &  &  &  & (7.44) & (9.06) & (7.38) \\
 $Rosz_2$ & 1000  & 1000  & 0. & 50 & 0.014 & 0.00118 & 114 &1.41  & 0.0341 & 0.593 \\
 &  &   &  & &  &  &  & (0.640) & (0.0295) & (0.597) \\
 \hline \hline
\end{tabular}
 \end{small}
\end{center}
\end{table}
\begin{table}[hbt]
\begin{center}
{ Table 2: Non-universal mSUGRA Models: Input  parameters are given at the
grand scale using  dimensions in GeV and the  convention $M_2>0$.
In all cases   $\mu>0$. }\\
  \vspace*{0.3cm}
\begin{small}
\begin{tabular}{||c|c|c|c|c|c|c|c|c||}
\hline \hline
 & $AD(fg9)$ & $SPS6_1$ & $SPS6_2$ & $SPS6_3$
 & $CDG_{200}$ & $CDG_{24}$ & $CDG_{75}$ & $CDG_{OII}$     \\ \hline
 $M_1$ & 420 & 480 & 720 & 960& 800 & 160 & -400 & 424 \\
$M_2$ & 420 &300 & 450 & 600&  160& 960 & 240& 200 \\
$M_3$ & 420 &300 & 450 & 600&  80& -320 & 80& 40 \\
$m_0$ &600 &150 & 225 & 300 & 1400 & 1400 &1400 &1400 \\
$m_{H_u}$ &$600\sqrt{2}$ &150 & 225 & 300 & 1400 & 1400 &1400 &1400 \\
$A_0$ & 420 & 0 & 0 &0  &1000 & 1000 & 1000 & 1000 \\
$\tan\beta$ & 40 & 10 & 10 & 10  &50 & 50 & 50 & 50 \\
  \hline \hline
\multicolumn{9}{||c||}{$v_{ij}\sigma (\tchi^0_i \tchi^0_j \to \gamma \gamma)$
  at $v_{ij}/c \simeq 10^{-3}$ in units of    $10^{-29} cm^3 sec^{-1}$} \\
  \hline
$\tchi^0_1 \tchi^0_1$ & 3.07 &0.482  &0.0973  & 0.0767 & 112 &0.000444  &0.00214  &72.2 \\
$\tchi^0_1 \tchi^0_2$ & 0.0102  & 7.58  &1.36  &0.386  &7.56  &0.0339   &0.00488   &11.3  \\
$\tchi^0_2 \tchi^0_2$ & 14.0 & 115  & 133  & 133 & 1.64 & 9.66  & 106  &1.1  \\
 \hline \hline
\multicolumn{9}{||c||}{$v_{ij}\sigma (\tchi^0_i \tchi^0_j \to gg )$
  at $v_{ij}/c \simeq 10^{-3}$ in units of    $10^{-29} cm^3 sec^{-1}$} \\
  \hline
$\tchi^0_1 \tchi^0_1$ & 18.3 & 1.6  &0.672   &0.376   &1356  & 0.118  &2.80   &13.6  \\
$\tchi^0_1 \tchi^0_2$ & 1.34  &0.986   &0.514  & 0.427 & 19.8 & 47.6  &2.82   &0.425  \\
$\tchi^0_2 \tchi^0_2$ &1.19  & 5.61  & 12.7  & 3529 &4.77  & 7.72  &4.70   &0.0286  \\
\hline
\multicolumn{9}{||c||}{$v_{ij}\sigma (\tchi^0_i \tchi^0_j \to gg )$
  at $v_{ij}/c \simeq 0.5$ in units of    $10^{-29} cm^3 sec^{-1}$} \\
  \hline
$\tchi^0_1 \tchi^0_1$ & 20.0 & 2.00  &0.811   &0.455   &565  & 0.131  &2.84   &11.9  \\
$\tchi^0_1 \tchi^0_2$ & 1.39 & 1.62 & 1.42&2.39&19.0&68.5 &2.50 &0.407 \\
$\tchi^0_2 \tchi^0_2$ & 1.67&9.42 &2418&5.17&4.86&7.79&8.20&0.0291\\
\hline \hline
\end{tabular}
 \end{small}
\end{center}
\end{table}
\begin{table}[htb]
\begin{center}
{ Table 3: GMSB and AMSB models.  Input parameters are at the big  scale,
  and $v_{ij}\sigma (\tchi^0_i \tchi^0_j \to \gamma \gamma, ~gg)$
  predictions are  at $v_{ij}/c \simeq 10^{-3}$   ($v_{ij}/c \simeq 0.5$).  }\\
  \vspace*{0.3cm}
\begin{small}
\begin{tabular}{||c|c|c|c||c|c|c|c|c|c||}
\hline \hline
\multicolumn{4}{||c||}{Parameters (Dimensions in $\rm GeV$) }&
\multicolumn{6}{c||}{$v_{ij}\sigma (\tchi^0_i \tchi^0_j
\to \gamma \gamma, ~gg)$ in $10^{-29} cm^3 sec^{-1}$ } \\ \hline
\multicolumn{4}{||c||}{ $\mu>0$ } & \multicolumn{3}{|c|}{ $\gamma \gamma$ }
& \multicolumn{3}{|c||}{ $gg$ }  \\
\hline \hline
\multicolumn{10}{||c||}{GMSB models; $n_e=n_q=N_{mes}=3$ } \\
\hline
Model & $M_{mess}$ & $M_{SUSY}$ &  $\tan \beta$
& $ \tchi^0_1 \tchi^0_1 $ &  $ \tchi^0_1 \tchi^0_2  $
& $ \tchi^0_2 \tchi^0_2  $ & $ \tchi^0_1 \tchi^0_1   $
&  $ \tchi^0_1 \tchi^0_2  $ & $ \tchi^0_2 \tchi^0_2  $    \\ \hline
$SPS7_1$ & 80000  & 40000   & 15 & 2.12  & 0.225 & 36.6  & 2.01 & 48.6  &7.53  \\
 &   &    &  &   &  &   & (2.40) & (20.9)  &(6.87)   \\
$SPS7_2$ & 120000 & 60000   & 15 & 1.15  & 0.124   &25.5  &0.409  &7.34  &2.61  \\
&   &    &  &   &  &   & (0.456) & (5.03)  &(2.43)   \\
$SPS7_3$ & 160000  & 80000   & 15 & 0.686  &0.0545 & 17.4 &0.146  &2.87  &1.04  \\
&   &    &  &   &  &   & (0.180) & (2.14)  &(0.991)   \\
$SPS8_1$ & 200000 & 100000   & 15 & 0.416  & 0.121   &131  & 0.181 &0.994  &4.78  \\
&   &    &  &   &  &   & (0.192) & (0.947)  &(8.30)   \\
 $SPS8_2$  & 400000 & 200000   & 15& 0.119  &0.00834    & 121  &0.0299  &0.0457  &7.13  \\
&   &    &  &   &  &   & (0.0298) & (0.0445)  &(26.9)   \\
 \hline \hline
 \multicolumn{10}{||c||}{AMSB models  } \\
\hline
Model & $m_0$ & $m_{aux}$ &  $\tan \beta$
& $ \tchi^0_1 \tchi^0_1 $ &  $ \tchi^0_1 \tchi^0_2  $
& $ \tchi^0_2 \tchi^0_2  $ & $ \tchi^0_1 \tchi^0_1   $
&  $ \tchi^0_1 \tchi^0_2  $ & $ \tchi^0_2 \tchi^0_2  $    \\ \hline
$SPS9_1$ & 450 & 60000   & 10 & 170   &0.0355    &0.487  & 0.229 &0.0237  & 0.315 \\
&   &    &  &   &  &   & (0.233) & (0.0240)  &(0.336)   \\
$SPS9_2$ & 675 & 90000   & 10 & 148  & 0.0149   & 0.217 & 0.110 & 0.00364 &0.179  \\
&   &    &  &   &  &   & (0.112) & (0.00374)  &(0.185)   \\
$SPS9_3$ & 900 & 120000   & 10 & 135  & 0.00822   & 0.122 &0.0652  &0.000984  & 0.116 \\
&   &    &  &   &  &   & (0.0659) & (0.00102)  &(0.117)   \\
 \hline \hline
\end{tabular}
 \end{small}
\end{center}
\end{table}

\vspace{0.5cm}
\noindent
\underline{The $\chi^0_1\chi^0_1\to \gamma\gamma$ mode.}\\
In all    universal mSUGRA models of Table 1,
 $\tchi^0_1$ is predominantly a Bino,
which implies rather small values for $v\sigma$, in agreement
with \cite{nnDM1, nnDM2}.
The  smallest  $v \sigma$ values,  in the range of
$ 10^{-33}~ cm^3sec^{-1}$,  appear
for the $SPS2$ models characterized by very heavy sfermions and charginos.
Such  values  increase  to the $10^{-31}~ cm^3sec^{-1}$ level,
for the $Rosz_2$, $SPS4$ and $AD(fg5)_2$  models,
where sfermions and charginos have masses of a few hundreds  of
GeV; and they  are  further   enhanced to  a few times
$10^{-30}~ cm^3sec^{-1}$,   in models like  $SPS3_1$ or $SPS3_2$
 where  the lightest $\chi^0_1$ is close  to   $\tilde\tau_1$.
These   enhancements remain  nevertheless
moderate,  because
 the intermediate $(\chi^0_1\chi^0_1 \to \tau^- \tau^+)$-contribution
 to the $\chi^0_1\chi^0_1\to \gamma \gamma $ amplitudes
is depressed by the small $\tau$ mass.

 The  non-universal mSUGRA set of models (Table 2)
includes cases like $AD(fg9)$,  in which the
 $\tchi^0_1$ is predominantly of Higgsino-type,
 leading to   $v \sigma \sim 3 \times 10^{-29}~ cm^3 sec^{-1}$ \cite{nnDM1, nnDM2}.
Much larger values like $v \sigma \sim 10^{-27}~ cm^3sec^{-1}$
are reached  in  the
$CDG_{200}$ and $CDG_{OII}$ cases though,  in which
$\tchi^0_1$ is predominantly a Wino combined  with
an appreciable Higgsino component.
The fact that  $\tchi^+_1$ is rather close to  $\tchi^0_1$
in the   $AD(fg9)$, $CDG_{200}$ and $CDG_{OII}$ models,  also plays a role
in   enhancing $v \sigma\chi^0_1\chi^0_1\to \gamma \gamma$.

In all other cases of Table 2, $\tchi^0_1$ is mainly a Bino
and $v \sigma $ is small. Among them, worth mentioning may the
 the $CDG_{75}$ model, which displays the very rare property to have
 a CP-odd  $\tchi^0_1$.

When it happens that $m_{A_0}\simeq 2m_{\tchi^0_1}$,
then it is very important to correctly include the $A^0$ width,
which we have done by using the code HDECAY \cite{HDECAY}.
An example of  a case where the $A^0$ width is necessary, is $CDG_{200}$.

 In Table 3,  we present  the results for the GMSB models
 $(SPS7,~SPS8)$, and the AMSB models $SPS9$   \cite{Snowmass}.
In $(SPS7,~SPS8)$, $\tchi^0_1$ is mainly a Bino and
$v\sigma$ rather small compared to the result in the  $SPS9$ model  where
$\tchi^0_1$ is mainly a Wino.
In   $(SPS7, ~SPS8)$ there exists
a near degeneracy of  $\chi^0_1$
with  $\tilde\tau_1$, which  enhances the intrinsically small Bino result
up to the $10^{-29}~ cm^3sec^{-1}$ level.  In $SPS9$,
 which is characterized by a near
degeneracy of the Wino-like  $\tchi^0_1$ with
  $\tchi^+_1$, the enhancement is  much bigger, leading to
  $v \sigma \sim 10^{-27}~ cm^3sec^{-1}$.

\vspace{0.5cm}
\noindent
\underline{The $\tchi^0_1\tchi^0_1\to gg$ mode.}\\
In Tables 1,2,3 we  give results both, for $v/c\simeq 10^{-3}$ and
$v/c\simeq 0.5$.
As in the $\gamma\gamma$ case, the largest $v \sigma$ values appear
for a Wino-like $\tchi^0_1$,
smaller ones for Higgsinos, and very small values for the Bino case.
On top of this, further enhancements may occur when $\tchi^0_1$ is
 very close to  $\tilde t_1$;
 or when $m_{A^0}\simeq 2 m_{\tchi_1} $ where
 a considerable sensitivity on $v/c$ may  also be induced,
 (see \eg  $CDG_{200}$ in Table 2).

 As seen from  Tables 1,2,3,
 the $v\sigma$ rates vary
from $2 \times 10^{-31}$ to $7 \times 10^{-29} cm^3sec^{-1}$ for
the universal m-SUGRA models; from $10^{-30} $ to $10^{-28}
cm^3sec^{-1}$ for the non-universal m-SUGRA models, with the
exceptional case of $CDG_{200}$ giving   $\sim 10^{-26}
cm^3sec^{-1}$; and from $3\times 10^{-31} $ to $2\times 10^{-29}
cm^3sec^{-1}$ as we move from the AMSB towards the GMSB models of
Table 3.

\vspace{0.5cm}
\noindent
\underline{The $\tchi^0_1\tchi^0_2$ and $\tchi^0_2\tchi^0_2$
annihilations.}\\
In the $\gamma\gamma$ case,  $v\sigma(\tchi^0_1\tchi^0_2)$ is very
often  considerably  smaller than $v\sigma(\tchi^0_1\tchi^0_1)$,
while $v\sigma(\tchi^0_2\tchi^0_2)$ considerably larger and often
reaching the $10^{-27}~ cm^3sec^{-1}$ level. This  is usually
driven by the fact that $\tchi^0_2$ is mainly a Wino in all models
where $\tchi^0_1$ dominantly a Bino.
 The occasional degeneracy of
$\tchi^0_2$ with $\tchi^+_1$, further enhances this value. The
only models where the $\tchi^0_2\tchi^0_2$ rate is smaller than
the $\tchi^0_1\tchi^0_1$ one, are  $CDG_{200}$,
$CDG_{OII}$ and the AMSB ones, in which  the Wino component of
$\tchi^0_1$ is usually considerably larger than that of $\tchi^0_2$.

In the $gg$ case, the $\tchi^0_1\tchi^0_2$ cross sections
can be either larger or smaller
than the $\tchi^0_1\tchi^0_1$ ones, because of the varying
situations for the squark  and chargino contributions,
 and orthogonalities that tend to
appear in certain couplings. In contrast to this,  $\chi^0_2\chi^0_2$
is most often  larger than  $\tchi^0_1\tchi^0_1$,
but the relative magnitudes  are often not so large as
in the $\gamma \gamma $ case.
Rates like $v \sigma(\chi^0_2\chi^0_2) \sim 10^{-28}~ cm^3sec^{-1}$
are often realized. Exceptional cases like $SPS3_1$   and $SPS6_3$
 give  $10^{-26}~ cm^3sec^{-1}$ at $v/c\simeq 10^{-3}$, while  $SPS6_2$
 gives  a similar value  at $v/c \simeq 0.5$
 at least partly  due to $A^0$ resonance effects.

\vspace{0.5cm}
\noindent
{\it \underline{Final comments}.}~~\\
In conclusion, we confirm that  $v_{ij}\sigma(\chi^0_i\chi^0_j\to \gamma \gamma,~gg )$
annihilation at $v_{ij}/c \sim 10^{-3}$, can assume
a wide range of values.
This arises because  these processes  start at one loop, thereby being
sensitive  to many aspects of the sfermion, ino
and Higgs boson masses, and   the peculiar mass-coincidences
appearing in many MSSM cases.

The results should be useful  for Dark Matter searches
through  neutralino annihilation to sharp photons.
Roughly, the gamma-ray flux expected in such a case from a direction
at an angle $\psi$ with respect to the direction of  the
galactic center,  is given by \cite{DM-distribution}
\bq
\frac{d{\cal F}}{d\cos\psi} \simeq ( 10^{-12}{\rm cm^{-2} s^{-1}})~
\Big [\frac{v \sigma(\chi^0_1\chi^0_1\to \gamma \gamma}
 {10^{-29}{\rm cm^3 s^{-1}}}\Big ] J(\psi)~,
\eq
where $J(\psi)$ depends on the dark matter density distribution along
the line of sight. Its magnitude   is expected to be of
order one at high $\psi$; but it increases as
$\psi$ decreases, and in some models  it may even reach values  up to $10^5$
when looking towards the galactic center  \cite{DM-distribution}.
On the basis of this, it seems that  for
$v \sigma(  \chi^0_1\chi^0_1\to \gamma \gamma) \gsim
 10^{-29}{\rm cm^3 s^{-1}}$, the observation
  of a sharp photon line with an
energy $E_{\gamma}$ precisely corresponding to the mass of
the LSP $\tchi^0_1$, should be possible \cite{DM-obs}.
 The actual observation of such energetic photons
 would supply   an interesting confirmation
of Supersymmetry, provided of course that the actual observations
would allow for acceptable values
of  the relic density of the lightest neutralino.
In this respect, the process $\chi^0_i\chi^0_j\to \gamma Z$
may also be interesting, and we are planning to extend our study to it.

The neutralino relic density
depends on the total annihilation cross section (summing all
final states like $f\bar f$, $WW$, $ZZ$, ...)\cite{DMrev}.
Concerning this,   we have found that the gluonic (sometimes even the photonic
 channel) may be important
in some models, and that occasionally considerable variations may appear in
$v_{ij}\sigma(\chi^0_i\chi^0_j\to gg )$, as $v_{ij}/c $
varies between $10^{-3}$ and $0.5$. In more detail,
the neutralino contribution to  cold dark matter (CDM) is
\bq
\Omega_{CDM} \simeq \frac{6 \cdot 10^{-27}{\rm cm^3 s^{-1}}}
{\langle \sigma_{\rm ann} v_{\rm rel} \rangle}
\eq
where $\langle \sigma_{\rm ann} v_{\rm rel} \rangle$ is the average
total neutralino annihilation cross section multiplied by the
relative velocity at $v_{\rm rel}/c \simeq 0.5 $. For a
model to be consistent with cosmological constraints,
we expect that
$\langle \sigma_{\rm ann} v_{\rm rel} \rangle \sim 10^{-26}{\rm cm^3 s^{-1}}$.
Thus, we note \eg from Table 2 that in the $CDG_{200}$ model,
 both $v \sigma (\tchi_1 \tchi_1\to gg)$ and
 $v \sigma (\tchi_1 \tchi_1\to \gamma \gamma )$ are of the order
 $10^{-27}{\rm cm^3 s^{-1}}$. A similar situation appears for
$v \sigma (\tchi_1 \tchi_1\to \gamma \gamma )$ in the
AMSB Snowmass models \cite{Snowmass} in Table 3.
We are tempted to conclude therefore that it would be safer,
if both these processes are
included in the relic density computations.

Because of the  sensitivity of the results  to the
SUSY models, and in order to contribute
in the successive stages of the required  analysis, we have
started the construction of a series of numerical codes
(PLATON) which will be made public so that anyone can
run his preferred set of MSSM parameters or explore the parameter
space.
At present the codes PLATONdml allow to compute  the annihilation rates
 $(v_{ij}/c) \sigma( \tchi^0_i\tchi^0_j\to\gamma\gamma)$
 at $v/c\simeq 10^{-3}$, while PLATONdmg and PLATONgrel
determine $(v_{ij}/c) \sigma (\tchi^0_i\tchi^0_j\to gg)$
at $v/c\simeq 10^{-3}$ and $v/c\simeq 0.5$ respectively, always in fb \cite{plato}.

If the emerging  overall picture  turns out
to be  cosmologically consistent,
it will  give a stringent test on SUSY models,
particularly if $\tchi^0_1\tchi^0_1\to \gamma Z $
is also observed with the correct properties.
 Particle Physics experiments should then provide further tests
by studying  neutralino production at high energy colliders,
through the inverse
processes $\gamma\gamma \to \tchi^0_i\tchi^0_j$ at LC and
$gg \to \tchi^0_i\tchi^0_j$ at LHC.
A theoretical study of these processes using the same
benchmark models as in the present paper is in preparation \cite{nn2}. \\

\noindent
Acknowledgments:\\
It is a pleasure for us to thank Abdelhak Djouadi for his
invaluable suggestions and help, and Jean-Lo\"ic  Kneur for
 very helpful discussions.


\end{document}